\documentstyle[12pt,aasms4]{article}
\newcommand\vev[1]{\left\langle #1\right\rangle}
\begin{document}
\title{Arc Statistics in Clusters: Galaxy Contribution}
 
\author{Ricardo A. Flores}
\affil{Department of Physics and Astronomy,
University of Missouri--St.Louis\\
St.Louis, MO 63121-4499\\
Electronic mail: Ricardo.Flores@umsl.edu}
 
\author{Ariyeh H. Maller}
\affil{Physics Department,
University of California--Santa Cruz\\
Santa Cruz, CA 95064\\
Electronic mail: maller@physics.ucsc.edu}

\and
 
\author{Joel R. Primack}
\affil{Physics Department,
University of California--Santa Cruz\\
Santa Cruz, CA 95064\\
Electronic mail: joel@ucolick.org}
 
 
\begin{abstract}
The frequency with which background galaxies appear as long arcs as a result
of gravitational lensing by foreground clusters of galaxies has recently
been found to be a very sensitive probe of cosmological models by Bartelmann
et al. (1998). They have found that such arcs would be expected far less
frequently than observed (by an order of magnitude) in the currently favored
model for the universe, with a large cosmological constant $\Omega_\Lambda
\sim 0.7$. Here we analyze whether including the effect of cluster galaxies on
the likelihood of clusters to generate long-arc images of background galaxies
can change the statistics. Taking into account a variety of constraints on the
properties of cluster galaxies, we find that there are not enough sufficiently 
massive galaxies in a cluster for them to significantly enhance the cross
section of clusters to generate long arcs. We find that cluster galaxies 
typically enhance the cross section by only $\lesssim 15\%$.
\end{abstract}


\section{Introduction}

The discovery of long arcs in clusters of galaxies (\cite{LP86};
\cite{SMFP87})
offered the prospect of using their observed frequency as a tool to test
cosmological models, using the paradigm of the frequency of quasar lensing
studies set forth by Turner, Ostriker, \& Gott (1984). Wu \& Mao (1996) were
the first to carry out
such a study, in order to gauge the influence of a cosmological constant on
the observed frequency of arcs in a homogeneous sample of EMSS clusters
(\cite{Letal94}, \cite{Letal99}). The main conclusion of Wu \& Mao (1996) was
that
in a spatially flat, low-density universe ($\Omega_m = 0.3$) one would observe
about twice as many arcs as in an Einstein-de Sitter universe, but still
only about a half the observed number of arcs. The discrepancy with the
observed number was somewhat larger though: observational restrictions reduce
the number considerably (\cite{HWY97}), but source evolution increases the
number (\cite{HF97}). A more recent study by Cooray (1999) found the number to
be in agreement with observations for a low-density universe (open or flat),
if a minimum cluster velocity dispersion of $\sim 1080$ km/s is assumed.
At least 5 of the eight clusters with confirmed arcs, though, have dispersions
below this (see \cite{Letal99}; and references therein). In these studies a
given cosmological model enters mostly via the geometry of space-time.

A recent study by Bartelmann et al. (1998), however, finds the
predicted number of arcs to be rather sensitive to the differences in
the properties of the clusters predicted in different cosmological
models. This study has sharpened the conflict between predictions and
observations of long arcs for the low-density flat CDM model.  They
find that an $\Omega_m=0.3$ open cold dark matter (OCDM) model
produces about as many arcs as are observed, but a spatially flat
$\Omega_m=0.3$ CDM model ($\Lambda$CDM) produces an order of magnitude
fewer, and a standard CDM (SCDM) model two orders of magnitude
fewer. Unlike previous studies, the difference in formation epoch and
concentration between clusters in different cosmological models was
consistently taken into account, and found to be mainly responsible
for the drastic differences in their predicted numbers of long arcs.
With many independent pieces of evidence indicating that
$\Lambda$CDM is the only concordant cosmological model (e.g., Bahcall
et al. 1999), it is rather surprising that such a model fails so
drastically the arc number test. Clearly, a close examination of
possible sources of uncertainty is warranted.

One possible source of enhancement in the observed number of arcs is the
contribution of the cluster galaxies to the creation of giant arcs. Previous
studies (\cite{WM96}; Hattori et al. 1997; \cite{HF97}; \cite{C99}) have
treated clusters as smooth mass distributions. 
The clusters in the dissipationless N-body simulations studied by
Bartelmann et al. (1998) have significant substructure, but they
could not resolve galaxies.  It is therefore desirable to
study the magnitude of the effect galaxies would have on the arc abundance
in clusters. Including galaxies in cluster lensing studies is not new (see
e.g. \cite{GN88}). Their effect in deep, high-resolution studies of individual
clusters, e.g. in A2218 (\cite{Ketal96}), AC114 (\cite{Netal98}) and A370
(\cite{Betal99}), has been found to be significant indeed. Here we quantify
their effect on arc statistics by calculating the ratio, $1+\Delta$, of the
cross section to produce long arcs\footnote[1]{
\ We concentrate only on arcs with length-to-width ratio
$\geq 10$ and length $\geq 8\arcsec$, the criteria of the search in X-ray
clusters (\cite{Letal99}).}
when cluster
galaxies are included to the cross section when they are not. Of course, the
comparison is to be made while keeping the projected mass in the field of
view fixed. We find that the results for $\Delta$ are surprisingly small,
typically less than 15\% (i.e. $\Delta \lesssim 0.15$), although there is
considerable scatter. The scatter can easily be reduced by averaging over 10
or so clusters.

We describe the gravitational lensing model we have used to calculate the
arc cross section in section \ref{model}. We then explore the observational
constraints on the various parameters of the model in section \ref{paramaters}.
Finally, we presents our results in section \ref{results} and conclude with
a discussion of our results and the conclusions we draw from them in section
\ref{discussion}, where we also comment on the recent work of Meneghetti et
al. (1999) on the same problem.
We use throughout a Hubble constant $H_o = 100 h$ km/s/Mpc.

\section{Gravitational Lensing Model} \label{model}

We model the main cluster mass distribution (dark matter plus the hot
intracluster gas) using the standard Pseudo Isothermal Elliptical Mass
Distribution (PIEMD), for which the bending angle components at position
$(x,y)$ relative to the cluster center are given by (\cite{KK93}; \cite{KK98})
\begin{eqnarray}
\theta_x = {b \over (1-q^2)^{1/2}} \tan^{-1}
\left[{(1-q^2)^{1/2} x \over \psi+r_{core}} \right]\ , \\
\theta_y = {b \over (1-q^2)^{1/2}} \tanh^{-1}
\left[{(1-q^2)^{1/2} y \over \psi+q^2 r_{core}}\right]\ ,
\end{eqnarray}
where $b = 4\pi(\sigma_{cl}/c)^2(D_{LS}/D_{OS})(e/\sin^{-1}e)$ and
$\psi^2 = q^2(x^2+r_{core}^2) + y^2$. The cluster has line-of-sight velocity
dispersion $\sigma_{cl}$ and core radius $r_{core}$. Its mass distribution has
intrinsic and projected axial ratios $q_3$ and $q$ respectively, and
$e=(1-q_3^2)^{1/2}$.
$D_{OS}$ and $D_{LS}$ are angular-diameter distances, from the observer to
the source and from the lens to the source respectively.

We model galaxies as truncated isothermal spheres (see \cite{Ketal96}), so
that the contribution to the bending angle at $(x,y)$ of a galaxy at
$(x_g,y_g)$ is given by
\begin{eqnarray}
\theta_x^g = b_g \left({r_{cut} \over r_{cut}-r_{core}^g}\right)
\left({x_g-x \over d}\right) \left[{(d^2+r_{core}^{g\ 2})^{1/2}-r_{core}^g
\over d} - {(d^2+r_{cut}^2)^{1/2}-r_{cut} \over d}\right]\ , \\
\theta_y^g = b_g \left({r_{cut} \over r_{cut}-r_{core}^g}\right)
\left({y_g-y \over d}\right) \left[{(d^2+r_{core}^{g\ 2})^{1/2}-r_{core}^g
\over d} - {(d^2+r_{cut}^2)^{1/2}-r_{cut} \over d}\right]\ ,
\end{eqnarray}
where $b_g = 4\pi(\sigma_g/c)^2(D_{LS}/D_{OS})$ and
$d^2 = (x_g-x)^2+(y_g-x)^2$. The galaxy has line-of-sight velocity dispersion
$\sigma_g$, core radius $r_{core}^g$, and truncation radius $r_{cut}$.

Figure 1 shows results for a fiducial cluster. The values of the various
parameters of the model are explained and justified in section
\ref{paramaters}, and summarized in Table 1. The cluster is at redshift
$z_{cl} = 0.2$, seen edge-on with $q = 0.75$, and has $\sigma_{cl} = 1200$
km/s and $r_{core} = 1 h^{-1}$ kpc. An Einstein-de Sitter universe and the
filled-beam approximation are assumed in the calculation of angular diameter
distances. The background cosmological model is not of great importance,
since we are here interested in the {\it difference} in the lensing cross
section of clusters to produce long arcs due to the inclusion of galaxies in
the clusters.

The top-left panel shows results for a smooth cluster (no galaxies). The
shaded area is the region behind which a circular source at redshift
$z_S = 1$, and of angular radius $0.5 \arcsec$, would be imaged into a long
arc farther out in the cluster. The inner (outer) dashed line is the
tangential (radial) caustic. The top-right panel shows the same results when
the galaxies are taken into account. The total mass inside a
$150\arcsec \times 150\arcsec$ field of view centered on the cluster,
shown in the bottom panels, has been kept fixed. It can be seen that there
is a significant distortion of the tangential caustic, which results in an
increased area where the source can be imaged into a long arc. This is shown
for the source marked as a filled star (at $x = 13\arcsec$ and
$y = 2\arcsec$), whose image can be seen in the bottom right panel. Note
there that the counter arc, marked by the arrow, would not
be seen given the typical magnitude of a long arc. Also, most arcs that
appear only when galaxies are taken into account are not formed on top of
galaxies, as noted early on by Grossman \& Narayan (1988). The circles mark
the positions
of the galaxies and have radii chosen to roughly correspond to the size of
the galaxies in a deep image.

The caustics labeled
1-4 in the top right panel are due to the correspondingly labeled galaxies
in the bottom right panel. The bottom left panel shows the critical curves
corresponding to the caustics in the top left (top right) panel as dashed
(solid) lines. The outer (tangential) dashed critical curve of the smooth
cluster is repeated in the bottom right panel. In general, the galaxies that
most distort and enlarge the shaded region in the top left panel are galaxies
close to this critical curve.

\section{Model Parameters} \label{paramaters}

In order to study the properties of images created by the cluster lens model
we need to specify all the parameters needed. Here we explain our choices
based on what is known about clusters and cluster galaxies.

The sample of clusters searched for long arcs is selected by X-ray flux
(strictly speaking, by central surface brightness; see the discussion in
\cite{Letal99}). This is expected to select very massive clusters given the
known correlation of X-ray luminosity with $\sigma_{cl}$
($L_x \sim \sigma_{cl}^4$) first established by Solinger \& Tucker (1972).
None of the EMSS
clusters with $L_x < 4\times10^{44}$ erg/s show any arcs (\cite{Letal99}),
corresponding to
a minimum dispersion $\sigma_{cl} = 784^{+68}_{-62}$ km/s using the recent
analysis of Wu, Xue, \& Fang (1999). Indeed, the lowest velocity dispersion
of the
clusters with arcs is $\sigma_{cl} \sim 800$ km/s (see \cite{Letal99}; and
references therein). Therefore we consider here clusters with
$\sigma_{cl} \ge 800$ km/s.

The shape of clusters is not known observationally.
Lens models that use the projected axial ratio $q_{BCG}$
of their brightest cluster
galaxy (BCG) reproduce rather well the orientation of arcs and arclets in
deep, high-resolution studies of clusters (e.g. \cite{Ketal96},
\cite{Netal98}). B\'{e}zecourt et al. (1999) find a somewhat rounder mass
distribution
to give a better fit. To the extent that $q_{BCG}$ is a good guide to $q$,
the study of Porter, Schneider, \& Hoessel (1991) implies $q \gtrsim 0.6$.
Numerical simulations
of clusters find triaxial shapes for galaxy clusters (\cite{Tetal99}), with
mean minor/major axial ratio of 0.5 for a low-density universe. It is easy
to translate their distribution into the distribution for $q$ assuming
nearly oblate or prolate halos (see \cite{B78}), from which we estimate that
$q \sim 0.5-0.9$ for most halos, with a median $q \sim 0.7$. We use
$q = 0.5, 0.75\ {\rm and}\ 0.9$ as representative values. This range covers
the values used in the studies of A370, A2218 and AC114.

Clusters are expected to have density profiles well approximated by the
Navarro, Frenk, \& White (1997) [NFW] profile,
$\rho \propto R^{-1}(R+R_s)^{-2}$. However,
in the radial range of interest here
it is the inner profile, where the density distribution changes from
$\rho \sim R^{-1}$ to $\rho \sim R^{-2}$, that really matters. It has been
argued (\cite{WNB99}) that a cluster with NFW profile cannot reproduce
the angular distance of arcs from their cluster centers (for the
dispersions of interest here, $\sigma_{cl} \sim 800-1400$ km/s), and the steep
inner profile of a BCG is needed. On the other hand, lensing studies of
several clusters
find that a core radius  $r_{core} \sim 30 h^{-1}$ kpc is needed
(e.g. \cite{TKD98}, \cite{Setal96}).
Here we use isothermal spheres with 
$r_{core} = 1\ {\rm or}\ 30 h^{-1}$ kpc to bracket
these results. A pure NFW profile would give results intermediate between
these two cases.

There are several parameters that describe the galaxies. First, we use
$r_{core}^g = 0.1 h^{-1}$ kpc throughout, in agreement with the constraints
from quasar lensing studies (see e.g. \cite{K96}). Second, we follow
standard practice in lensing studies (see e.g. the discussion in \cite{K96})
and assign a velocity dispersion to a galaxy using a Faber-Jackson
(\cite{FJ76}) relation $\sigma_g = \sigma_*(L/L_*)^{1/\beta}$. The luminosity
$L$ is drawn from a Schechter distribution,
$(L/L_*)^{\alpha}\ {\rm exp}(-L/L_*)$
(\cite{S76}). The value of $\beta$ ranges from $\beta \sim 3$ in the B band
to $\beta \sim 4$ in the infrared (\cite{dVO82}). Here we use $\beta = 3$;
our results do not change much if we use $\beta = 4$ instead, as we discuss
below.

Our next step is to choose the galaxy truncation radius, $r_{cut}$.
The truncation
of galaxy halos inside clusters has been studied numerically by
Klypin et al. (1999). An estimate of the size of a halo at distance $R$ from a
cluster center is given by the tidal radius $r_t$. For a galaxy and cluster
with $r_{core} = 0$, $r_t = (\sigma_g/\sigma_{cl})R$. Since we do not know the
distance $R$ for a galaxy at projected separation $r$ from the cluster center,
we use the average distance along the line of sight,
$\vev{R} = \int{R(r,z)\rho(r,z)dz}/\int{\rho(r,z)dz}$. The galaxies that
contribute the most to the arc cross section are those near the
critical curve of the smooth cluster (see Fig. 1). Therefore, we evaluate
$\vev{R}$ at the Einstein radius,
$r \sim 60h^{-1}(\sigma_{cl}/1200\ {\rm kms}^{-1})^2$ kpc. For $\sigma_{cl}
\sim 1200$ km/s, this is comparable to $R_s$ for such a cluster (see
\cite{Tetal99}). Thus, we evaluate $\vev{R}$ at projected separation $r=R_s$
and obtain $\vev{R} \sim 2r$ for the NFW profile. Finally,
we compare the rotation curve of a numerical halo (the typical example
discussed by \cite{Ketal99}, bottom curve of their Fig. 6) to the model
rotation curve of a truncated isothermal sphere, and obtain $r_{cut} \sim
3r_t/4$.

We shall take here $\vev{R} = 100 h^{-1}$ kpc, and use throughout the paper
$r_{cut} = 3r_t/4$. Therefore,
$r_{cut}^*=15(\sigma_*/230\ {\rm kms}^{-1})(\sigma_{cl}/1200\ {\rm kms}^{-1})$
$h^{-1}$ kpc. It is interesting to note that this value agrees fairly well
with the value inferred from the effect of galaxies on the spatial
distribution and orientation of the arcs and arclets in the cluster A2218
(see \cite{Ketal96}). The scaling of $r_{cut}$ with $\sigma_g$ implies
that for a given cluster
$r_{cut} = r_{cut}^*(L/L_*)^{\gamma}$ with $\gamma = 1/\beta = 1/3$.
Thus, the scalings of $\sigma_g$ and $r_{cut}$ with $L$ 
are different from
those suggested by Brainerd, Blandford, \& Smail (1996) ($\beta=4$,
$\gamma=1/2$), and
used in the studies of A370, A2218 and AC114. However, we find the arc cross
section to be very similar in either case because the galaxy mass-to-light
ratio, $M/L \propto \sigma_g^2 r_{cut}/L$, is constant in both cases. Some
authors have also explored $\gamma=0.8$ based on studies of the fundamental
plane of ellipticals that suggest $M/L \propto L^{0.3}$ (see \cite{Netal98},
and \cite{Betal99}). It has been noted, however, that the fundamental plane
can also be interpreted assuming constant $M/L$ (\cite{BBF92}), so the jury
is still out on this question. Finally, we also
note that we expect $r_{cut}^* \propto \sigma_{cl}$ to be smaller for
lower-dispersion clusters. This might be part of the reason for the different
$r_{cut}^*$ obtained in the AC114 and A2218 analyses (see \cite{Netal98} and
\cite{Ketal96}, respectively).

In order to find the effect of galaxies on the arc cross section we must also
choose their characteristic velocity dispersion $\sigma_*$, how they are
distributed inside a cluster, and how many there are. Since galaxies too faint
and/or too far from the critical curve do not contribute much to the arc cross
section, we find it enough to include galaxies down to 2 mag fainter than $L_*$
inside an area corresponding to $150\arcsec \times 150\arcsec$ for a cluster at
redshift $z_{cl} = 0.2$ (see Fig. 1). This will be our fiducial field of view
(FOV) at this redshift, and we shall study a region of the same physical size
at other redshifts. We are interested in clusters in the redshift range
$z_{cl} = 0.2-0.6$, the range in which the arc cross section is large (see
\cite{Betal98}).

Smail et al. (1998) have studied a sample of 10 clusters at $z_{cl} \sim 0.2$
with
X-ray luminosities in the range of interest here. They find that the surface
number density of red galaxies is $\propto r^{\delta}$, with
$\delta = -0.96 \pm 0.08$, and that the luminosity distribution is well fit
by a Schechter function with $\alpha = -1.25$ and
$M_V^* = (-20.8 \pm 0.1) + 5{\rm log}h$. Using $M_V^* = -20.8$ and
$M_V = -20.35 - 8.5({\rm log}\sigma_g - 2.3) + 5{\rm log}h$ (\cite{dVO82}),
we infer $\sigma_* = 226$ km/s. From their Table 2 we infer a count
of $20-40$ galaxies (down to 2.3
mag fainter than $L_*$) in a $150\arcsec \times 150\arcsec$ FOV, with a
mean of 32. Furthermore, Smail et al. (1997) have studied a set of 10 clusters
in the redshift range $z_{cl} = 0.37-0.56$. They find that elliptical
galaxies are distributed with $\delta = -0.8 \pm 0.1$ in the radial range of
interest here, and their luminosities are well described by a Schechter
function with $\alpha = -1.25$. Furthermore, we derive a count of $29$
galaxies per cluster in a $150\arcsec \times 150\arcsec$ FOV for
their clusters at $z_{cl} \sim 0.4$ (down to 2 mag fainter than $L_*$),
consistent with a count of $20$ at $z_{cl} = 0.2$ assuming equal numbers
in areas of equal physical size. We find that the same holds, within errors,
for their clusters at $z_{cl} \sim 0.54$.

The previous results also agree with a homogeneous sample of clusters
at low redshift (\cite{Letal97}). We infer a count of $17$ galaxies in our
FOV at $z_{cl} = 0.2$ for the mean of the sample (down to 2 mag fainter
than $L_*$, assuming $\delta = -1$ and equal numbers in equal areas).
However, most of these systems have low velocity dispersions. For the
only cluster with $\sigma_{cl} \sim 1200$ km/s, the fit parameters imply
$30$ galaxies. The mean $M_{b_j}^*=-20.2$ implies $\sigma_* = 232$ km/s,
assuming
a mean color $b_j-V \sim .7$. Finally, we also infer similar counts from the
detailed study of 7 rich Abell clusters at $z_{cl} \sim 0.15$ by
Driver, Couch, \& Phillipps (1998).
They find $\sim 50-150$ galaxies inside a $280 h^{-1}$ kpc radius (down to 3
mag fainter than $L_*$; see their fig. 11). Thus, we infer $19-57$ galaxies
in our FOV, with a mean of 38 (down to 2 mag fainter than $L_*$, using their
fig. 6 and assuming both $\delta = -1$ and equal numbers in equal areas).

We shall summarize these observations by adopting  $\sigma_* = 230$ km/s and
$\alpha = -1.25$. We shall assume a universal luminosity function, an
adequate assumption for the inner region of rich clusters (see \cite{DCP98}).
Finally, we shall take $\delta = -1$.\footnote[2]{
\ Strictly speaking, we distribute the galaxies in projection just like the
cluster surface density profile, including flattening and core radius.}
Therefore, the galaxies
trace the dark matter, in agreement with gravitational lensing studies of
clusters (see \cite{TKD98}; and references therein).

The choice of the number of galaxies in our FOV is complicated by the fact
that it depends on the cluster velocity dispersion (\cite{B81}).
Girardi et al. (1999) have computed total cluster luminosities within fixed
physical
radii for a large, homogeneous sample of 89 clusters for which there is also
velocity dispersion data. We have analyzed their data for luminosities inside
$0.5 h^{-1}$ Mpc, and find that the data are well fit by a cluster luminosity
$L_{cl} \sim 6.3(\sigma_{cl}/770\ {\rm kms}^{-1})^{1.5} \times 10^{11} h^{-2}
L_{\odot}$. There is, of course, significant scatter. This is believed to be
physical, and results from the fundamental plane of clusters (\cite{Setal93})
seen in projection. We find that 68\% of the clusters have luminosities
$(.67-1.5)L_{cl}$. Extrapolating the validity of $L_{cl}$ to
$\sigma_{cl} = 1200$ km/s, we infer that there should be $37$ galaxies in
our fiducial FOV (assuming $\delta = -1$ and equal numbers in equal areas).
In view of this, and the previous discussion, we shall take $N_g = 40$
galaxies inside a square $316 h^{-1}$ kpc on a side (our FOV at
$z_{cl} = 0.2$) for a cluster with $\sigma_{cl} = 1200$ km/s, but we explore
the range $N_g = 20-60$ as representative of the likely scatter to be
encountered. For clusters with different $\sigma_{cl}$ we scale $N_g$ by
$(\sigma_{cl}/1200\ {\rm kms}^{-1})^{1.5}$.

We finish this discussion of our choice of parameters by summarizing
them in Table 1.

\section{Results} \label{results}

We have calculated $\Delta$ for Monte Carlo realizations of the galaxy
distribution in a cluster by ray tracing through a fine grid in the image
plane (we find that $0.375\arcsec$ spacing works well enough) to find the
points that are imaged back to a given source. We take circular sources
(this is adequate for our purpose of finding the cross section for very long
arcs, for which the intrinsic ellipticity of the sources is not important)
of $1\arcsec$ diameter, and at redshift $z_S = 1$. Sources are placed with
$0.25\arcsec$ spacing or smaller depending on $z_{cl}$. Sets of contiguous
pixels in the image plane that trace back to a given source are then an image.
If at least one image has angular area at least 10 times the area of the
source, the pixel area around the source position is added to the arc cross
section.

Our main results are presented in Table 2, where we give the average $\Delta$
of 10 realizations of the distribution of galaxies in a cluster,
$\vev{\Delta}$. Results are
given for a given cluster at three different redshifts and for three
representative axial ratios
$q = q_3$ (i.e. the cluster is seen edge-on; see discussion below).
The top row at each redshift gives results
for a cluster with $\sigma_{cl} = 1000$ km/s and $r_{core} = 1(30) h^{-1}$
kpc, while the bottom row gives results for $\sigma_{cl} = 1200$ km/s and
$r_{core} = 1(30) h^{-1}$ kpc. The sources are asumed to be at redshift
$z_S = 1$. Based on the scatter of 100 realizations of the galaxy
distribution in a cluster at $z_{cl}=0.2$, with $q=0.75$ and
$\sigma_{cl} = 1200$ km/s, we estimate the
error for $\vev{\Delta}$ in Table 2 to be $\pm 0.02$.

Thus, we see that typically $\vev{\Delta} \lesssim 0.12$. Increasing the
number of galaxies to $N_g = 60$ changes $\vev{\Delta}$ only to
$\vev{\Delta} = 0.15$ from $\vev{\Delta} = 0.11$ for our fiducial cluster
at $z_{cl} = 0.2$:
$\sigma_{cl} = 1200$ km/s, $r_{core} = 1 h^{-1}$ kpc, and $q=0.75$. Also,
for the entire range $\sigma_{cl} = 800-1400$ km/s we find that
$\vev{\Delta} = 0.07-0.15$ for the same cluster. The scatter introduced by
the discrete nature of galaxies (the numerical scatter introduced by the
finite size of our grids on the source and image planes is very small)
is such that in 68\% of the realizations $\Delta$ is in the range
$\Delta = 0.03-0.16$, again for our fiducial cluster. The results are not
sensitive to the assumed source redshift. For $z_S = 1.2$ or $0.8$,
$\vev{\Delta} \sim .08$ for the same cluster. We do not find our neglect
of the dependence
of $r_{cut}$ on the distance of a galaxy to the cluster center to be important
either. We have used $r_{cut} = (\sigma_g/\sigma_{cl})\vev{R}$, where
$\vev{R} = (\pi/2)r$ for a galaxy at projected separation r if we assume
$\rho \propto R^{-3}$, as appropriate for the more distant galaxies. We find
that this changes $\vev{\Delta}$ only to $\vev{\Delta} = 0.13$ from
$\vev{\Delta} = 0.11$ for our fiducial cluster.

Our numbers are given for edge-on clusters for simplicity. However, the
projection effect could significantly increase $\vev{\Delta}$ only for fairly
flattened clusters seen nearly face-on. For example, for a cluster at
$z_{cl} = 0.2$ with $\sigma_{cl} = 1200$ km/s and $r_{core} = 1 h^{-1}$ kpc,
$\vev{\Delta} = 0.037$ if $q_3 = q = 0.9$ (see Table 2). We find that this
changes to $\vev{\Delta} = 0.067$ instead if the cluster has $q_3 = 0.5$, and
is seen in projection with $q = 0.9$. If we considered the cluster to be
prolate instead, $\vev{\Delta}$ would be smaller. Thus, our results for
$\vev{\Delta}$ are not significantly different when projection effects are
taken into account.

We have assumed a Schechter luminosity function throughout. This often
underestimates the number of bright galaxies in a cluster (see e.g.
\cite{Letal97}). We have corrected for this by assuming a luminosity function
$\propto (L/L_*)^{\alpha}\ {\rm exp}(-(L/L_*)^{1/4})$. We find that this
functional form (with the same $\alpha$) fits the data better for $L > L_*$,
without changing much the galaxy count for $L < L_*$. However, we find that
with this luminosity function the value of $\vev{\Delta}$ changes only to
$\vev{\Delta} = 0.12$ from $\vev{\Delta} = 0.11$ for our fiducial cluster.

A possible concern is that these
results apply only to a smooth cluster, whereas the clusters in the
simulations are clearly substructured. However, we have done a realization
of a substructured cluster by adding a large, secondary clump away from
the center of the cluster, and described by a truncated isothermal sphere
density profile with $\sigma_{cl} = 500$ km/s and $r_{core} = 1 h^{-1}$ kpc.
In this case we took $r_{cut} = 225h^{-1}$ kpc $\sim 2r$, where r is the
projected separation. This is clearly large given its velocity dispersion,
but we took this value to maximize the effect of this subclump in the
calculation. We find that even in this case, keeping the
same total mass in our FOV as in the case of a smooth cluster
with $\sigma_{cl} = 1200$ km/s, $\vev{\Delta}=0.074$ instead of
$\vev{\Delta}=0.11$.

\section{Discussion \& Conclusions} \label{discussion}

Our main conclusion from this study is that the likelihood that a cluster
generate long-arc images of background sources is not significantly enhanced
by the presence of its galaxies. The many observationally based constraints
that we have taken into account imply that there are simply not enough
sufficiently massive galaxies in a cluster to affect significantly the
probability of a long arc. The effect could be more significant for the
probability of finding arcs of certain characteristics. For example, typically
the long arcs appear isolated and aligned more or less perpendicular to the
cluster major axis (\cite{Letal99}). It can be seen in Figure 1 that the cross
section for
those arcs (the shaded area outside the left and right side of the tangential
caustic in the top left panel) is enhanced more by the presence of the
galaxies: $\Delta \sim 0.4$. The effect would also be much larger for
arclet statistics, which we have not addressed here.

Undoubtedly out treatment is simplified, but it is clear that the presence of
galaxies within a cluster is a minor effect that cannot reconcile the observed
frequency of arcs in clusters with the expectations in a universe dominated
by a cosmological constant.

Meneghetti et al. (1999) have recently studied this problem with a different
methodology. Our studies are fairly complementary. For example, their clusters
have realistic large-scale substructure, whereas we explore more systematically
the galaxy distribution parameter space. Our results are consistent; e.g. their
ensamble of clusters with galaxies generate about 7\% fewer long arcs than
their pure dark matter clusters, a result entirely within the range we find
here. Our results make it clear that the effect of galaxies is not necessarily
to decrease the number of arcs, and that the number can be significantly larger
in individual clusters.

Acknowledgments: One of us (RF) would like to acknowledge the organizers
of the first Princeton-PUC Workshops on Astrophysics (held in Pucon, Chile,
January 11-14, 1999) for their invitation to present our results prior to
publication.  AHM and JRP acknowledge support from NASA and NSF grants
at UCSC.


\clearpage

{\scriptsize
\begin{deluxetable}{lccccccccccc}
\tablenum{1}
\tablecaption{Model Parameters (Fiducial parameters marked with $^{\ddagger}$)}
\tablehead{
\colhead{Parameter} & \colhead{$\sigma_{cl}$} & \colhead{$r_{core}$} &
\colhead{$q$} & \colhead{$\sigma_*$} & \colhead{$r_{cut}^*$} &
\colhead{$r_{core}^g$} & \colhead{$N_g$} & \colhead{$\alpha$} &
\colhead{$\beta$} & \colhead{$\gamma$} & \colhead{$\delta$}\\
\cline{2-12} \\
\colhead{} & \colhead{km/s} &\colhead{$h^{-1}$ kpc}& \colhead{} &
\colhead{km/s} & \colhead{$h^{-1}$ kpc} & \colhead{$h^{-1}$ kpc} &
\colhead{} & \colhead{} & \colhead{} & \colhead{} & \colhead{}}
\startdata
Value & 1000, 1200$^{\ddagger}$ & 1$^{\ddagger}$, 30 &
0.5, 0.75$^{\ddagger}$, 0.9 & 230 & 15 & 0.1 & 40$^{\ddagger}$ &
-1.25 & 3 & 1/3 & -1\nl
\enddata
\end{deluxetable}
}

{\scriptsize
\begin{deluxetable}{lccc}
\tablenum{2}
\tablecaption{Average Galaxy Contribution to Arc Cross Section, $\vev{\Delta}$}
\tablehead{
\colhead{} & \multicolumn{3}{c}{axial ratio $q$}\\
\cline{2-4}\\
\colhead{redshift $z_{cl}$} & \colhead{0.5} & \colhead{0.75} & \colhead{0.9}}
\startdata
0.2 &  0.029(0.11)   &  0.060(0.009)  & -0.001(0.017) \nl
    &  0.063(0.095)  &  0.11(0.12)    &  0.037(0.031) \nl
0.4 & -0.052(-0.069) &  0.047(0.025)  &  0.035(0.079) \nl
    &  0.028(0.019)  &  0.11(0.071)   &  0.069(0.047) \nl
0.6 & -0.095(-0.094) & -0.042(-0.077) & -0.029(-0.061)\nl
    & -0.025(-0.04)  &  0.019(0.038)  &  0.039(0.061) \nl
\enddata
\end{deluxetable}
}

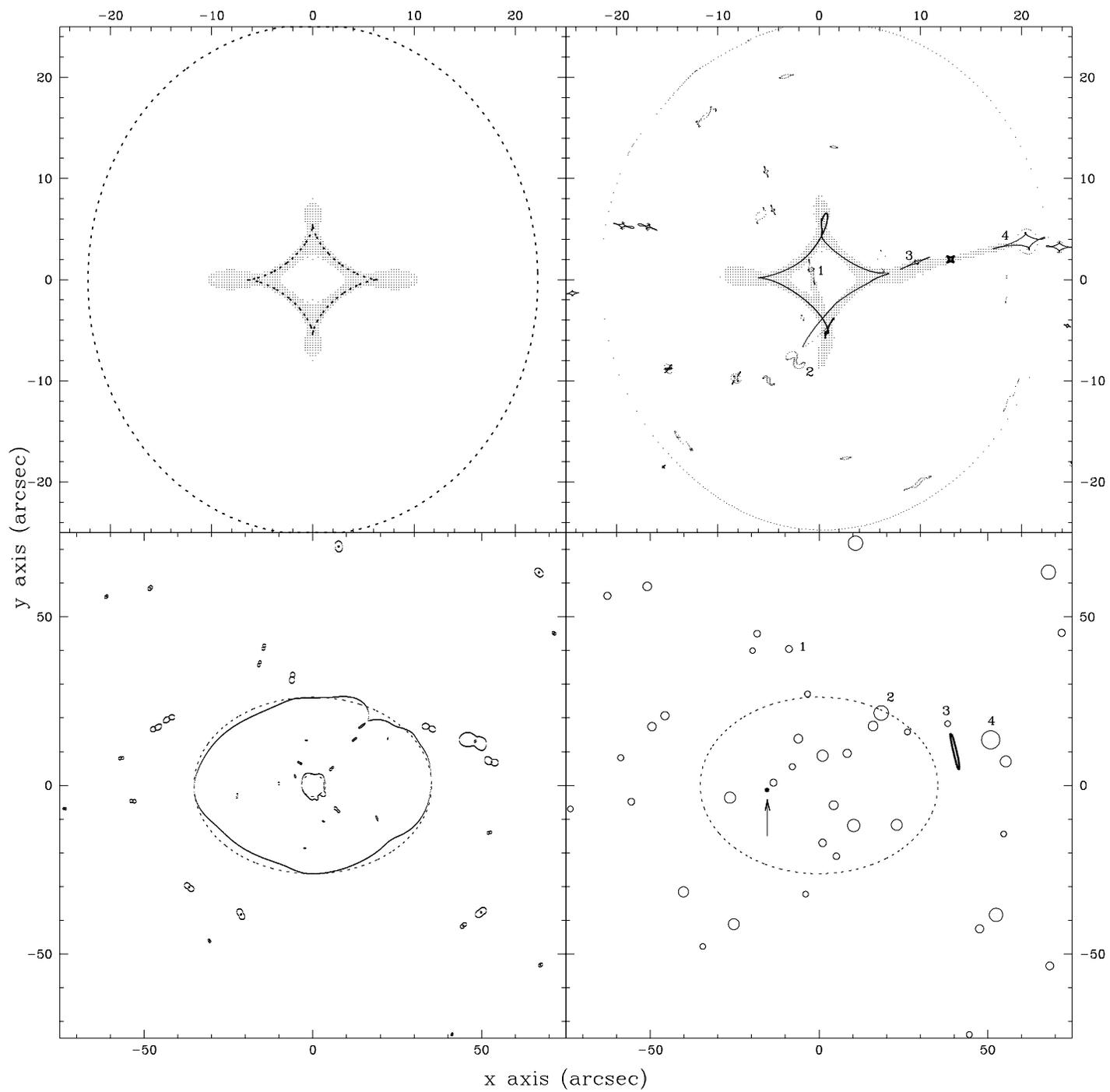
\begin{figure}
 
\caption{Results for our fiducial $\sigma_{cl} = 1200$ km/s cluster. Top
panels: source plane. Bottom panels: image plane. See text for further
explanation.}
 
\end{figure}


\end{document}